\documentclass[graybox, envcountchap]{svmult}

\usepackage{mathptmx}        
\usepackage{amsmath}
\usepackage{amssymb}
\usepackage{color}
\usepackage{helvet}          
\usepackage{courier}         
\usepackage{dirtree}

\usepackage{makeidx}        
\usepackage{graphicx}        
\usepackage{subfig}

\usepackage{multicol}        
\usepackage[bottom]{footmisc}

\usepackage{hyperref}        
\hypersetup{colorlinks=true,urlcolor=blue,citecolor=blue}
\usepackage[numbers]{natbib}
\usepackage[misc]{ifsym}

\makeindex             

\begin{document}


\title{Astrophysical and Cosmological Searches for Lorentz Invariance Violation}
\author{Shantanu Desai}
\institute{Shantanu Desai (\Letter) \at Dept of Physics, IIT Hyderabad, Kandi, Telangana, 502284 \email{shntn05@gmail.com}}

%
%
\maketitle

\abstract{Lorentz invariance is one of the fundamental tenets of Special Relativity, and has been extensively tested with laboratory and astrophysical observations. However, many   quantum gravity models  and theories beyond the Standard Model of Particle Physics predict a violation of Lorentz invariance at energies close to the Planck scale. This article reviews  observational  and experimental tests  of Lorentz invariance violation (LIV) with  photons, neutrinos and gravitational waves. Most astrophysical tests of LIV using photons are  based on searching for a correlation of the spectral lag data with redshift and energy. These have been primarily carried out using compact objects such as pulsars,  Active Galactic Nuclei (AGN), and Gamma-ray bursts (GRB). There have also been some claims for LIV from some of these spectral lag observations with GRBs, which however are in conflict with the most stringent limits obtained from other LIV searches. Searches have also been  carried out using polarization measurements from GRBs and AGNs.
For neutrinos, tests have been made using  both astrophysical observations at MeV energies  (from SN 1987A) as well as  in the TeV-PeV energy range based on  IceCube observations,  atmospheric neutrinos, and long-baseline neutrino oscillation experiments. Cosmological tests of LIV entail looking for a constancy of the speed of light as a function of redshift using  multiple observational probes, as well as looking for birefringence in  Cosmic Microwave Background observations. This article will  review all  of these aforementioned observational tests of LIV, including results which are in conflict with each other.}


\section{Introduction}
\label{sec:1}

In Special Relativity, the speed of light, $c$, is a Lorentz invariant quantity, which is same in all inertial frames of reference, due to the fact that laws of Physics are the same in all observers in an inertial frame of reference. However, this 
\emph{ansatz} is not valid in many theories beyond the Standard Model of Particle Physics and also various quantum gravity and string theory models.   
In these models, Lorentz invariance is expected to be broken at ultra-high energies close to the Planck scale ($E_{pl} \sim 10^{19}$ GeV), and
the speed of light is consequently a function of photon energy~\cite{GAA98}. Alternately,   one could think of  the vacuum refractive index as been  different from unity in such models~\cite{Ellis}. The energy-dependent speed of light can then  be written as~\cite{AmelinoCamelia98}:
\begin{equation}
    v(E) = c\left[1 - s_{\pm}\frac{n+1}{2} \left(\frac{E}{E_{QG}}\right)^n\right],
    \label{eq:vE}
\end{equation}
where $s_{\pm} = \pm 1$ denotes the sign of the Lorentz Invariance violation (LIV), corresponding to subluminal ($s_{\pm}=+1$) or superluminal ($s_{\pm}=-1$); $E_{QG}$ denotes the quantum gravity scale where LIV effects kick in, and $n$ is a model-dependent term and is usually equal to one or two, corresponding to linear or quadratic LIV, respectively.  The correspondence between $n$ and  different theories of LIV  is discussed in ~\cite{Addazi}.

A plethora of searches for LIV  have been carried out using photons, neutrinos, and gravitational waves. The astrophysical sources used for searches of LIV with photons include pulsars, Active Galactic Nuclei (AGN), Gamma-Ray Bursts (GRBs), and Cosmic Microwave Background (CMB). Most of these searches using AGNs and GRBs use spectral lags, 
which can be  defined as the difference in the  arrival times  between high energy and low energy photons, and is positive if the high energy photons arrive earlier than the low energy ones~\cite{Norris96}.
For a source at cosmological distances, Equation~\ref{eq:vE} leads to a first-order differential time delay for signals
as a function of energy~\cite{Jacob}:
\begin{equation}
\frac{\partial t}{\partial E} \approx \frac{1}{H_0 E_{QG}} \int_0^z \frac{(1+z')dz'}{h(z')},
\label{eq:deltdelE}
\end{equation}    
where $h(z) \equiv \frac{H(z)}{H_0}$ is the dimensionless  Hubble parameter as a function of redshift. For the current standard $\Lambda$CDM model~\cite{Planck18}, $h(z)= \sqrt{\Omega_M (1+z^\prime)^3 + \Omega_\Lambda}$. The expression for $h(z)$ in other cosmological models can be found in ~\cite{Biesiada}. 

The observed spectral time lag ($\Delta t_{obs}$)  corresponding to an energy interval $\Delta E$ can be modeled as a sum of two independent delays: 
\begin{equation} 
\Delta t_{obs} = t_l -t_h = \Delta t_{int}^{obs} + \Delta t_{LIV}, 
\label{eq:sumdeltat}
\end{equation}
where $\Delta t_{obs}$ is the observed spectral lag; $t_h$ and $t_l$ represent the arrival times of photons of energies
$E_h$,  $E_l$ where $E_h>E_l$;   $\Delta t_{int}^{obs}$ is the intrinsic time delay due to astrophysical emission in the observer frame, and $\Delta t_{LIV}$ due to LIV.

The intrinsic time delay in the observer frame is given by 
\begin{equation}
 \Delta t_{int}^{obs} =(1+z)  \Delta t_{int},
 \label{eq:nullhypothesis}
\end{equation}
where the $(1+z)$ term accounts for the cosmological time dilation. 
The time delay due to the linear and quadratic LIV models can be obtained from $n=1$ and $n=2$, respectively of the following equation~\cite{Jacob},
\begin{equation}
\Delta t_{LIV}  = -s_{\pm}\left(\frac{1+n}{2H_0}\right)\left(\frac{E_h^n - E_l^n}{E_{QG,n}^n}\right)\int_{0}^{z} \frac{(1+z^{\prime})^n}{h(z^{\prime})} \, dz^{\prime}, 
\label{eq:deltaliv}
\end{equation} 
where $E_{QG,n}$ is the Lorentz-violating or quantum gravity scale, above which Lorentz violation becomes the dominant contribution, and $E_h$, $E_l$ represent the upper and lower energy intervals for photons  arriving at times $t_h$ and $t_l$, respectively.
In Eq.~\ref{eq:deltaliv}, $n=1$ and  $n=2$ denote  linear  and  quadratic LIV models, respectively. 
We should point that the time delay  in Eq.~\ref{eq:deltaliv} corresponds to one explicit model of LIV~\cite{Jacob}, which assumes that the spacetime translations are not affected in any LIV model. Alternative LIV and doubly-special relativity models have also been proposed where quantum gravity affects spacetime translations~\cite{Rosati}. Therefore, for such models, Eq.~\ref{eq:deltaliv} would not be valid. Here, we shall mostly discuss results which use the formulation in ~\cite{Jacob}.  We also note that all results prior to the work in ~\cite{Jacob} had a missing $(1+z)$ term in the integrand of Eq.~\ref{eq:deltaliv}. The reason that is that earlier derivations equated the proper distances the photons of different energies traversed when they reached Earth.  However, the source and the Earth have fixed comoving distance between them, but the proper distance changes with the expansion of the universe~\cite{Jacob}. Therefore, one must take into account the expansion of the universe, while considering the relative delays  between particles at different energies~\cite{Jacob}.  Results on constraints on  Lorentz violating Standard Model Extension (SME)~\cite{Kostelecky} from astrophysical observations (circa 2008) can also be found in ~\cite{Kostelecky08}. We also note that many works report limit on photon mass obtained from dispersion relations while searching for LIV. Other techniques include looking for signatures of photon splitting and photon decay~\cite{WeiWu21}.


This chapter will collate all the observational results on LIV using the aforementioned spectral lag method as well as various other astrophysical objects and observations such as pulsars, polarization observations (from AGNs, GRBs, CMB), cosmology, neutrinos, and also a few terrestrial experiments.  We should also mention that many other reviews focusing on observational tests of LIV have also been written over the years. A non-exhaustive list includes ~\cite{Sarkar02, Bolmont11,Ellis13,Bolmont20,Horns,Patrick, Stecker17,Lang20,Wei2021,WeiWu21,Wei22,HeMa22,Carlos,Lang22} (see also references therein).  However, since this is a rapidly evolving field, this review will attempt to  cover a few topics not mentioned in some of the aforementioned reviews.
Corresponding reviews  emphasizing the  theoretical aspects of LIV  can be found in ~\cite{Jacobsen,Tasson1,Tasson2,Liberati,Maccione09, Mattingly,GAA,Wei2021,HeMa22,Rosati,Addazi} and will not be discussed  here. A review of experimental tests of special relativity, which is the theoretical foundation behind  Lorentz invariance can be found in ~\cite{Will05,Altschul}. A compilation of limits on photon mass obtained  using pulsars, fast radio bursts, solar wind, solar system planets, etc. can be found in Particle Data Group~\cite{PDG} as well as ~\cite{Wei2021} (See Table 3 therein) and are not reviewed here. Limits on LIV from gravitational Cherenkov radiation are discussed in ~\cite{Tasson15}. This review will also not discuss the closely related topic of  tests of Weak Equivalence Principle using astrophysical observations  (for eg.~\cite{Boran}) or tests of variation of fundamental constants~\cite{Bora,Srinikitha} (also Chapter 13 of this book).

The outline of this chapter is as follows. We shall review LIV searches using pulsars in Sect.~\ref{sec:pulsars}; Active Galactic Nuclei in Sect.~\ref{sec:AGN}; neutrinos in Sect.~\ref{sec:neutrinos}; cosmological observations in Sect.~\ref{sec:Cosmology}; gravitational waves in Sect.~\ref{sec:GW}. LIV searches with GRBs (which occupy bulk of this chapter) are discussed in Sect.~\ref{sec:GRBs}. Searches with polarization measurements are reviewed  in Sect.~\ref{sec:polarization}. We conclude in Sect.~\ref{sec:conclusions}.

\section{Pulsars and other Galactic sources}
\label{sec:pulsars}
Pulsars are rotating neutron stars, which emit pulsed radio emissions   with periods from  milliseconds to a few seconds  and  magnetic fields ranging from $10^8$ to $10^{14}$ G~\cite{handbook,Reddy}. Pulsars have proven to be   wonderful  laboratories for tests and applications of a whole range  of topics in Physics and Astronomy~\cite{Blandford92,Kahya}.

The first test of LIV using pulsars was carried out using EGRET observations of the Crab pulsar from 70 MeV to 2 GeV~\cite{Kaaret}. This work looked an energy dependent pulsar arrival time. No statistically significant variation of the pulse arrival time with energy was found. Subsequently, a 95\% lower limit on the energy scale of quantum gravity $E_{QG}$ of $1.8 \times 10^{15}$ GeV was set~\cite{Kaaret}. A similar search was subsequently  carried out from the Crab pulsar in the energy range 600 GeV$-$1.2 TeV by the MAGIC collaboration~\cite{Magiccrab}. Since no such energy-dependent variation was found, upper limits were set for both superluminal ($s_{\pm}=-1$) and subluminal cases ($s_{\pm}=1$). These limits are $E_{QG}>(4.5-5.5) \times 10^{17}$ GeV for linear LIV and $E_{QG}>(5.3-5.9) \times  10^{10}$ GeV for quadratic LIV. 
 Recently, sub-PeV  observations of the Crab nebula from the Tibet~\cite{TibetAS} and HAWC~\cite{HAWC} collaborations have also been used to set lower limits on LIV scale by looking for photon decay and photon splitting~\cite{Satunin}. Based on the absence of photon decay  and splitting, lower limits of $E_{LIV}>4.1 \times 10^{14}$ GeV and $E_{LIV}>1.9 \times 10^{13}$ GeV, respectively were set~\cite{Satunin}. The HAWC collaboration also did a similar analysis of the Crab pulsar and three other  pulsars (J1825-134, J1907+063, J0534+220, J2019+368) to set limits on LIV with lower limits of (100$-$200) $E_{pl}$~\cite{HAWC2}.
The Crab pulsar has also been observed at over 1 PeV by the LHAASO-KM2a collaboration with the most energetic photon having an energy of $1.12 \pm 0.09$ PeV~\cite{LHAASOCrab}. 

Due to LIV, a free electron can emit Cherenkov radiation in vacuum. Assuming the inverse Compton spectrum of the Crab nebula exceeds up to PeV energies, a constraint on LIV was obtained based on  the stability of $\geq$ 1 PeV electrons against electron Cherenkov radiation. Thus,  a lower limit on $E_{QG} > 10^{25}$ GeV was obtained~\cite{LiMa22}.

In 2021, the LHAASO collaboration detected 12 gamma-ray sources with energies from 100 TeV up to  1.4 PeV with a statistical significance of $>7\sigma$~\cite{LHAASO12}. Except for the Crab pulsar, these sources have not been confirmed from previously  known point sources. A search for superluminal LIV was done using this dataset by looking for single photon pair production as well as  single photon splitting into multiple photons~\cite{LHAASO2}. Both these processes lead to a cut-off in the spectrum. The LIV parameters were obtained using a forward modelling procedure. No spectral cutoff due to LIV was seen and therefore  a lower limit on LIV was set. The 95\%  confidence level (c.l.) lower limit on first order LIV is $E_{QG} > 10^5 E_{pl}$~\cite{LHAASO2}. The limits on second order LIV is given by $E_{QG} > 10^{-3} E_{pl}$~\cite{LHAASO2}. An independent search for superluminal LIV of this data was also carried out~\cite{LiMaLHAASO}, which obtained approximately similar lower limits as in ~\cite{LHAASO2}. Furthermore, it was also pointed out that the observation of the highest energy photon (1.4 PeV) could be compatible with subluminal LIV with a scale of $3.6 \times 10^{17}$ GeV~\cite{LiMaLHAASO}.

A search for LIV  in the context of quantum electrodynamics was also done using the Crab pulsar~\cite{Stecker01}. In this setup, the maximum vacuum speed for an electron ($c_e$) is different from $c$ and can be written as: 
\begin{equation}
c_e \equiv c_{\gamma} (1 \pm \delta),
\label{eq:ce}
\end{equation}
where  $|\delta|<<1$. For $\delta<0$, photon decays to an electron-positron pair for photons with energies $E_{max}=m_e\sqrt{2/|\delta|}$. From the observations of Crab pulsar at 50 TeV~\cite{Tanimori98}, we get $\delta>-2 \times 10^{-16}$~\cite{Stecker01}. If $\delta>0$, electrons become superluminal for energies $E>0.5 E_{max}$ and will emit radiation at all frequencies due to Cherenkov radation. Based on the observations of electrons  in primary cosmic-ray with energies up-to 1 TeV~\cite{Nishimura80}, one gets $E_{max}\geq 2$ TeV implying $\delta<1.3 \times 10^{-13}$~\cite{Stecker01}.

\section{Active galactic nuclei}
\label{sec:AGN}
All tests of LIV using AGNs have been carried out using observations of blazars at TeV gamma rays  with atmospheric Cherenkov telescopes. Blazars are AGNs with jets pointed directly at us and with Lorentz factors of about 10~\cite{Urry}.
The key idea in looking for LIV is to look for correlation of spectral lag with  energy  to constrain $E_{QG}$  using Eq.~\ref{eq:deltaliv}.
The first test of LIV using AGN observations was made based on the TeV gamma-ray flare from Mrk 421 on 15 May 1996 detected by the Whipple atmospheric Cherenkov telescope~\cite{Biller}. By comparing the arrival times of photons with energies  $<$ 1 TeV and $> 1$ TeV, a 95\% c.l. lower limit of $E_{QG}> 0.4 \times 10^{17}$ GeV was set. Subsequently, the MAGIC collaboration analyzed two TeV flares of Markarian 501 (located at $z=0.034$) between May and July 2005  in the energy range from 0.15 - 10 TeV. Based on the time delay, a lower limit of $E_{QG}>0.21 \times 10^{18}$ GeV and $> 0.26 \times 10^{11}$ GeV for linear and quadratic LIV, respectively was obtained~\cite{MAGIC08}. Note that these limits are valid for both subluminal and superluminal propagation.  The HESS Collaboration then did a similar analysis looking for photon time delays using the TeV gamma-ray flare observed from  PKS 2155-304 on July 28, 2006. Two different  methods based on wavelet and the method of cross-correlation  were used to estimate the time delays between the light curves of different energies. This method then yielded 95\% c.l. lower limits of $E_{QG}>(5.2-7.2) \times 10^{17}$ GeV for linear LIV and $E_{QG}>1.4 \times 10^{9}$ GeV for quadratic LIV, which are valid for both superluminal and subluminal effects~\cite{Hess08}. This result was subsequently superseded by a refined analysis technique consisting of an event by event likelihood fit (applied to individual photon events) as applied to the flare data and new lower limits of $E_{QG}>2.1 \times 10^{18}$ GeV and $E_{QG}> 6.4  \times 10^{10}$ GeV at 95\% c.l. were obtained~\cite{Hess11}.

The HESS collaboration also used observations of  a VHE flare with $E>100$ GeV from  PG 1553+113 (located at a redshift $z=0.49 \pm 0.05$) observed on the  night of April 26 and 27, 2012 to constrain LIV~\cite{Abra15}. No energy dependent delay in the photon arrival time was seen. Consequently, a lower limit of $E_{QG}>4.11 \times 10^{17}$ GeV and $E_{QG}>2.1 \times 10^{10}$ GeV was set for linear and quadratic LIV, respectively, applicable for both superluminal and subluminal propagation~\cite{Abra15}.

An independent analysis of the above flare has also been carried out~\cite{Shao10}.  This work obtained lower limits of $E_{QG}>4.9 \times 10^{16}$ GeV, $1.2 \times 10^{17}$ GeV, and $2.6\times 10^{18}$ GeV for Mrk 421, Mrk 501,  and PKS 2155-304, respectively for  linear LIV~\cite{Shao10}. For a quadratic model, the corresponding limits obtained were $E_{QG}>1.5 \times 10^{10}$ GeV and $9.1\times 10^{10}$ GeV for Markarian 421   and PKS 2155-304, respectively~\cite{Shao10}.

The HESS collaboration detected another TeV flare from Mrk 501 in June 2014  with multi-TeV variability on time scales of minutes with its  energy spectrum extending up to 20 TeV~\cite{HESS19}. This data was used to search for evidence of LIV for both $s_{\pm} =+1$ as well as $s_{\pm} =-1$. The best-fit lower limit for linear LIV model  is given by  $E_{QG}> 2.6,3.6 (\times 10^{17})$ GeV for $s_{\pm}=1$ and $s_{\pm}=-1$, respectively~\cite{HESS19}. The corresponding limits for quadratic  LIV model are given by
$E_{QG}> 8.5,7.3 (\times 10^{10})$ GeV for $s_{\pm}=1$ and $s_{\pm}=-1$ respectively~\cite{HESS19}.

Some of the earlier limits on LIV obtained  using Mrk 421, Mrk 501, and PKS 2155-304 were later disputed in ~\cite{LiMaAGN}. This work showed that with a careful re-analysis of the light curves, the earlier upper limits should be revised and are consistent with light speed variation as a function of energy, which were hinted at from GRB observations (to be discussed in the subsequent section). More details on these methods and comparison with the previous limits on LIV for the same 
object can be found in ~\cite{LiMaAGN}.

In addition to the aforementioned searches for LIV from single sources, limits have also been set using a stacked catalog of blazars  based on the distortion in the observed spectra, which gets accumulated during propagation. Two examples of such an approach include Ref.~\cite{Biteau} and Ref.~\cite{Lang19}. The key idea  behind this is that, in the presence of LIV, the threshold for pair production gets lowered~\cite{Jacob}, which affects the optical depth of  photons with energies larger  than  10 TeV  from extragalactic sources. In the first such study, 86 spectra from 38 AGNs were used~\cite{Biteau}. A joint fit to the absorption signature seen in gamma-rays using models of extragalactic background light (EBL) along with $E_{QG}$ was done.  The first work which applied this technique obtained a 95\% c.l. lower limit on $E_{QG}$ given by $E_{QG} > 0.6 E_{pl}$ for linear LIV and $s_{\pm}=1$~\cite{Biteau}.  This  method of  stacking spectra from multiple AGNs was then improved  by selecting the optimum bins to account for the uncertainty in extragalactic background light~\cite{Lang19}. They applied this technique to 18 measured spectra from six AGNs detected at TeV energies from HESS, HEGRA, VERITAS, and TACTIC for sub-luminal propagation (corresponding to $s_{\pm}=1$). The 95\% c.l. lower limit on $E_{QG}$ for linear and quadratic LIV are given by $E_{QG}>12.8 \times 10^{19}$ GeV and $E_{QG}>2.38 \times 10^{12}$ GeV, respectively~\cite{Lang19}. These constitute  some of the most stringent limits on LIV to date.

A search for LIV within QED was also carried out with observations of Mrk 501 (along the same lines as described in Sect.~\ref{sec:pulsars} for Crab pulsar), based on the fact that there is no decrease in the inferred optical depth based on the observed spectrum~\cite{Stecker01}. This work obtained a limit on $\delta$ given by $\delta\leq 2 \times 10^{-16}$~\cite{Stecker01}.

\section{Neutrinos}
\label{sec:neutrinos}
Limits on LIV from neutrinos have been obtained from both neutrino oscillation experiments based on atmospheric neutrinos, (where neutrinos created with a specific lepton flavor are observed at the detector in a different lepton flavor),  as well as using  astrophysical neutrino observations,  and also using direct tests of speed of neutrinos using long baseline neutrino oscillation experiments. We briefly summarize some of these key results. Other reviews of tests of Lorentz invariance and CPT violation  using neutrinos can be found in ~\cite{Diaz14,Diaz}.

The year 1998 was a watershed moment in the history of Physics  when evidence for neutrino oscillations was found using the Super-Kamiokande experiment from atmospheric neutrinos~\cite{Messier}, and subsequently re-affirmed with more statistics and  using all the combined atmospheric neutrino samples~\cite{SK04,Desai04,SKshowering}. After the first evidence for atmospheric neutrino oscillations~\cite{Messier},  LIV was  also considered as a possible mechanism for these  neutrino oscillations~\cite{Coleman98}. For oscillations due to LIV, the oscillation probability scales as $L \times E$ as opposed to $L/E$ for atmospheric neutrino oscillations~\cite{Fogli98}, where $E$ is the neutrino energy and $L$ is the total neutrino pathlength. However the scaling of o the scillation probability with energy  using the 1998 data was found to be    $L \times E^{-0.9 \pm 0.4}$, when $L$ is normalized to 1000 km and $E$ is expressed in GeV~\cite{Fogli98}. This indicates  that the  oscillations seen in Super-K  cannot be due to LIV. In 2015, the Super-K collaboration carried out a dedicated search for LIV using the full neutrino dataset from 100 MeV to over 100 TeV and did not find any evidence for the same~\cite{SuperKLIV}. Consequently, constraints on coefficients of Lorentz violating SME were set. A similar search was done using the IceCube neutrino dataset between 100 GeV and 1 PeV also provided null results~\cite{IceCubeLIV}.  

In 2011, there was a flurry of excitement when the Opera experiment claimed evidence for superluminal neutrinos. However about a year later, new sources of errors were found in the original measurements and the refined analysis showed that there is no evidence for supeluminal neutrinos~\cite{Opera11}. Other long baseline neutrino oscillation experiments such as ICARUS~\cite{ICARUS} and MINOS~\cite{MINOS}  have also  found no evidence for superluminal neutrinos. More details on the searches and  potential implications for superluminal particles can be found in a recent review~\cite{Ehrlich}.

The first (and to-date the only) observation of astrophysical MeV neutrinos from SN 1987A from the LMC  at a distance of 50 kpc opened up the era of multi-messenger astronomy with neutrinos~\cite{IMB,Kamiokande}.
It was shown from the simultaneous 
arrival of photons and neutrinos after a journey of 160,000 years, that the velocity of neutrinos is comparable to the speed of light with an accuracy of $\mathcal{O} (10^{-9})$~\cite{Longo87}. Similar constraints on the neutrino speed were obtained using  observations of PeV neutrinos from TXS 0506+056 by IceCube with difference between neutrino speed and speed of light equal to within $\mathcal{O} (10^{-12})$~\cite{Laha} (see also ~\cite{Boran18}), which is three orders of magnitude more stringent than SN 1987A. From the same event a constraint of $E_{QG}>3 \times 10^{16}$  GeV was obtained~\cite{Ellisneutrino,kaiwang}.

Searches for LIV using astrophysical neutrinos have also been done using neutrinos with energies in the TeV-PeV range. Following the first detection of two 1 PeV cascade events by the IceCube collaboration~\cite{IceCube13},  a lower limit of $E_{QG}>10^{-5} E_{pl}$ and 
 $E_{QG}>10^{-4} E_{pl}$ was obtained for linear and quadratic LIV, respectively assuming both the events are extragalactic~\cite{Borriello}.
There have also been a class of searches which have set limits on LIV by looking for association between GRBs and IceCube neutrinos~\cite{AC2016,AC2,Huang,Huang2,ZhangYang22}. We provide some highlights from these aforementioned works. More details can be found in the original references. 
Huang et al~\cite{Huang} found 13 IceCube neutrinos with energies between 60 TeV and 2 PeV, which are associated with GRBs.
They showed that the arrival time of neutrino events is consistent with an energy-dependent speed of light variation indicating positive evidence for LIV with an energy scale of $E_{LIV} = (6.4 \times 1.5) \times 10^{17}$ GeV.
A search for  correlation between the high energy starting events in 7.5 years of IceCube data  and more than 7000
GRBs was done in ~\cite{ZhangYang22}. This work found six possible  GRB-neutrino pairs~\cite{ZhangYang22}. They used a variant of Eq.~\ref{eq:deltaliv} as follows~\cite{ZhangYang22}:
\begin{equation}
\Delta t =  \frac{1}{H_0} \frac{s_{\nu}E_{\nu}- s_p E_p}{E_{QG}}\int_{0}^{z} \frac{(1+z^{\prime}) dz^{\prime}}{\sqrt{\Omega_m (1+z^{\prime})^3+\Omega_{\Lambda}}}
\label{eq:deltalivnu}
\end{equation} 
where $z$ is the GRB redshift, $E_{\nu}$, $E_{p}$ are the neutrino and photon energies respectively; $s_{\nu}$ and $s_{p}$ are the sign factors for the photon and neutrino, respectively.

Based on this, a lower limit of $E_{QG} > 8 \times 10^{17}$ GeV was set~\cite{ZhangYang22}.
One issue with the above (and similar other results~\cite{AC2016,AC2}) is that no thorough statistical analysis to assess the significance of the association between GRBs and neutrinos has been done. It is entirely likely that the neutrinos could be due to atmospheric neutrino or diffuse astrophysical neutrino background (not associated with GRBs). The analysis done by the  IceCube collaboration does not find any evidence for neutrinos spatially or temporally correlated with GRBs~\cite{IceCubeGRB,IceCubeGRB2}.
Nevertheless, as IceCube is continuing to take data and with future upgrades such as IceCube-Gen2~\cite{IceCubeGen2} as well as PINGU~\cite{PINGU},  it may be possible to detect ultra-high energy neutrinos from GRBs, which would shed additional light on LIV.

\section{Cosmological observations}
\label{sec:Cosmology}
The standard ($\Lambda$CDM) cosmological model consisting of 70\% dark energy and 25\% dark matter agrees well with large scale structure observations~\cite{Planck18}.   Recently a number of observational tensions have been found in this standard model such as $H_0$ tension, $\sigma_8$ tension etc~\cite{Periv18}.  The edifice of the standard model is General  Relativity which subsumes Special Relativity.
Therefore, Lorentz Invariance and the constancy of speed of light  is inbuilt in the standard cosmological model. Although no one has thoroughly  explored the implications of LIV based theories to  resolve or alleviate some of the cosmological tensions, many tests of alternatives of $\Lambda$CDM, which posit a varying speed of light as a function of redshift ($c \equiv c(z)$) have been carried out. Some of these theories can solve the cosmic coincidence problem.
One such  class of models consist of  Varying speed of light theories~\cite{Moffat92,Barrow98}. Many tests of varying speed of light theories have been carried out using cosmological observations using   baryon acoustic oscillations (BAO), galaxy clusters, strongly lensed sources along with Type Ia SN data~\cite{Bengaly,Borac,LiuCao21,VSL}. The key idea behind testing the variability of speed of light is  to test for variation of $D_a (z)$:
\begin{equation}
    D_A(z) = \frac{1}{(1+z)}\int_0^z \frac{c dz}{H(z)},
    \label{eq:da}
\end{equation}
where $c$ is the speed of light. Equation~\ref{eq:da} can be generalized by replacing $c$ with a varying $c(z)$. A cosmological measurement of the variability of speed of light has been carried out using the Pantheon Type 1a Supernovae  sample, BAO measurements  and cosmic chronometer observations~\cite{Bengaly}. This work found values of $c$ at two different redshifts  given by $c=(3.2 \pm 0.16) \times 10^5$ km/sec at $z \approx 1.58$ and $c= (2.67 \pm 0.14) \times 10^5$ km/sec at $z \approx 1.36$. Although the measurement at $z \approx 1.58$ is consistent with $c$ to within the $1\sigma$,   the measurement at $z \approx 1.36$ points to a $2.3\sigma$ discrepancy with $c$. However, this estimate also depends on the details of the non-parametric reconstruction and some of the systematic errors have been underestimated~\cite{Bengaly}.  
A similar study with  galaxy clusters done by combining the measurements of galaxy cluster gas mass fractions, $H(z)$
from cosmic chronometers, and Type-Ia Supernovae
also found no variation of speed of light with redshift~\cite{Borac}. Another test for the  variation of the speed of light was done by combining data from four strongly lensed systems  (SLACS, BELLS, SL2S and LSD) along with Type Ia Supernovae~\cite{LiuCao21}. No deviation in the speed of light from $c$ as found was a function of redshift from this analysis~\cite{LiuCao21}.
Various other astrophysical and cosmological tests of varying speed of light theories are reviewed in ~\cite{VSL}.

Another consequence of LIV is the rotation of linear polarization also known as  birefringence, where the photon dispersion relation depends on the polarization. This causes a rotation of the CMB polarization and induces parity odd cross-correlations~\cite{Lue99}. If the Lagrangian contains a parity violating term, for example a Chern-Simons term of the form 
${\cal L}_{\rm CS}=-(1/2)p_{\alpha} A_{\beta}\tilde{F}^{\alpha\beta}$, where $\tilde{F}^{\alpha\beta}$ and  $A_{\beta}$ denote the  dual electromagnetic tensor and the vector potential, respectively, the Chern-Simons term would cause the two polarization states of photons propagate with different group velocities, which cause the polarization plane to rotate by an angle $\Delta \alpha$~\cite{Carroll90}. If we posit $p_{\alpha}$  to be the derivative of a scalar field $\phi$,  we get $\Delta \alpha =(\Delta \phi)/E$, where $E$ corresponds to an arbitrary energy scale~\cite{Komatsu09}.

The first such search for cosmic birefringence was done using the rotation measure limits from 5 years of WMAP data~\cite{Komatsu09}, and a limit on the effective photon mass was found with $m_{\gamma} \leq 3.8 \times 10^{-19} eV$ ~\cite{Durrer08}. Recently,  a $2.4\sigma$ evidence for non-zero birefringence  was found using the third Planck public release, with the birefringence angle ($\beta$) given by $\beta=0.35 \pm 0.14$~\cite{Minami18}.  A refined value of $\beta=0.30 \pm 0.11$ was obtained using  the fourth Planck public release, with the significance getting enhanced to 3$\sigma$~\cite{Diego22}.
Further implications of this result and prospects for confirmation with upcoming experiments are reviewed in ~\cite{Komatsu22}.

Other tests with CMB data  include signatures of  specific LIV theories such as effect of Einstein-Aether 
theory on the CMB anisotropy spectrum~\cite{Zuntz08} and birefringence due to Lorentz violating Standard model extension~\cite{Caloni22}.

\section{Gravitational Waves}
\label{sec:GW}
The year 2016  ushered in the era of gravitational wave  (GW) astronomy with the detection of binary black hole merger GW150914 by the  LIGO-VIRGO collaboration~\cite{LIGOBBH}. To-date the LIGO-VIRGO collaboration has detected GWs from about 100 binary black hole,  binary neutron star or black hole/neutron star binaries. Tests of LIV with gravitational waves entail a direct test of speed of gravitational waves.
A direct test of speed of gravitational waves using the observed time delay between the detection of the signal at the Hanford and Livingston detector sites  has been done based on  the first binary black hole merger (GW150914), which showed that $c_{gw}< 1.7 c$~\cite{Blas}. The near-simultaneous detection of GWs and photons from GW 170817 also enabled a direct test of relative vacuum velocity between GWs and photons $-3 \times 10^{-15} < \left(\frac{c_{gw}-c}{c}\right) < 7 \times 10^{-16}$~\cite{LIGOGRB}.
Other indirect tests based on Shapiro delay of GWs can be found in ~\cite{Boran,Kahya16}.

In addition to above model-independent tests of speed of GWs, constraints on LIV SME using GW observations have also been obtained.
Following the first detection of GW150914,
the effects of LIV on the dispersion and birefringence of GWs were calculated and a limit set using  GW150914~\cite{KosteleckyGW}. This analysis was then extended to the first full GW transient catalog released by the LIGO-VIRGO collaboration (GWTC-1)~\cite{ShaoGW,WangGW}. This analysis has recently  been extended to  GW events in both GWTC-1 and GWTC-2~\cite{WangShaoLiuGW,ZhaoCaoGW} as well as GWTC-3~\cite{ZhaoCaoGW,HaegelGW}.

\section{Gamma-Ray Bursts}
\label{sec:GRBs}
The vast majority of tests of LIV have been done using GRBs  for more than two decades starting with ~\cite{AmelinoCamelia98}, which did  the first  feasibility study demonstrating  the power of GRBs for  probing LIV.  Doing a thorough comprehensive summary of over two decades of literature is beyond the scope of this short review. So we shall mainly report  highlights each of the works which used GRBs for LIV searches. More details can be found in the original references or in some of the other observational reviews on LIV mentioned in the Introduction.

GRBs are single-shot explosive events, which have been observed over ten  decades in energies from  keV to  100~TeV range~\cite{Kumar}.  They are located at cosmological distances, although a distinct time-dilation signature in the light curves has not yet been demonstrated~\cite{Singh}. GRBs are traditionally divided into two categories based on their durations, with  long (short) GRBs lasting more (less) than two seconds~\cite{Kouveliotou}.
Long GRBs are usually associated with core-collapse SN~\cite{Woosley} and short GRBs with neutron star mergers~\cite{Nakar}. There are however many exceptions to the aforementioned dichotomy, and arguments  for additional GRB sub-classes have also been made~\cite{Kulkarni,Bhave} (and references therein). 
Similar to AGNs,  most searches of  LIV with GRBs have been done using  spectral lags.  Besides spectral lags, a few tests have also been done using polarization observations. In this section we discuss results from spectral lag searches starting with fixed observer frame energies and then using fixed energy intervals in the source frame. In Sect.~\ref{sec:polarization}, we shall  discuss LIV searches with  GRBs using polarization measurements. We note that a tabular summary of results for LIV using GRB spectral lags (upto 2021) has also been  collated in  Table 1 of ~\cite{WeiWu21}.

\subsection{Searches in fixed observer frame energies}
The first study for LIV using GRB spectral lags was carried out  using GRB 930229 based on the simultaneous arrival of a flare with a rise time of  $(220 \pm 30) \mu$-sec that occurred simultaneously from 30 $-$ 200 keV~\cite{Schaefer99}. From this GRB, a limit on fractional variation of speed of light given by $\Delta c/c < 6.3 \times 10^{-21}$  was set, where $\Delta c$ refers to any energy-dependent deviation in the speed of light~\cite{Schaefer99}. Subsequently, data from 6 GRBs with measured redshifts from the BATSE as well as the OSSE detector was analyzed to look for energy dependent speed of light by looking  for a correlation in the spectral lag with redshift and also stochastic variation in photon velocities at the same energies~\cite{Ellis99}. No such variation was found and subsequently a limit of $E_{QG} > 10^{15}$ GeV was set~\cite{Ellis99}. A refined analysis using wavelets  increased the lower limit to  $E_{QG} > 6.9 \times 10^{15}$ GeV (for subluminal propagation)~\cite{Ellis03}. The first test of LIV using GRB spectral lags in the MeV energy range was carried out using three different energy bins from 0.2-17 MeV using data from the RHESSI satellite~\cite{Boggs04}. No time lags were found. The limit on $E_{QG}$ was obtained  from Eq~\ref{eq:deltdelE}, assuming the source is at redshift of $z\approx 0.3$, although this was not a direct redshift estimate. This work obtained a limit of $E_{QG} > 1.8 \times 10^{17}$ GeV and $5.5 \times 10^7$ GeV for  linear and quadratic LIV, respectively~\cite{Boggs04}.

The first such systematic study by stacking a  GRB sample was carried out in ~\cite{Ellis,Ellis07}, who considered a sample of 35 GRBs in the redshift range $z=0.168-4.3$ from HETE, BATSE, and Neil Gehrels SWIFT. This was the first work where the spectral lags were modeled  as the sum of a constant intrinsic lag  along with another contribution due to   LIV (Eq.~\ref{eq:sumdeltat}). This was also the first work where an extra $(1+z)$ term was used inside the integrand in Eq.~\ref{eq:deltaliv} following the arguments in ~\cite{Jacob}. They also  found a $4\sigma$  evidence  for the  higher energy photons to arrive earlier than the lower energy ones, which at face value points to evidence for LIV with $s_{\pm}=-1$~\cite{Ellis}. However, when an additional systematic offset was added to enforce the $\chi^2$/DOF  to be equal to one,  the statistical significance for LIV reduced to about $1\sigma$. Subsequently,  a lower limit of $E_{QG} \geq (0.9-2.1) \times 10^{16}$ GeV was set at 95\% c.l~\cite{Ellis}. Note that even though at face value, the results showed tentative hints for $s_{\pm}=-1$, the limits in ~\cite{Ellis} were obtained for $s_{\pm}=1$.
The same data was also used to search for LIV by combining  it  with BAO and Union2 Type Ia SN data, and  within the aegis of  three cosmological models, namely $\Lambda$CDM, wCDM, and Chevallier-Polarski-Linder model~\cite{Pan15}. The results were consistent with a zero slope indicating that there is no evidence for LIV~\cite{Pan15}.
A joint cosmological and LIV analysis of the same data  along with JLA Type Ia SN data has also been carried out, which showed that the slope is also  consistent with no LIV to within 2$\sigma$~\cite{Zou}.

Subsequently, a similar test for spectral lag variation with energy was done using GRB 051221A, located at  $z=0.5465$, using light curves from  Swift-BAT, and Konus-Wind~\cite{Piran06}. For the Swift data, the spectral lags were estimated using four different energy bins 15-35 keV, 50-150 keV, 100-150 keV, and 300-350 keV. Based on the Swift time resolution of 4~msec, no time lags were detected between the peaks in the different energy bands and a limit on LIV energy scale was set with values given by $(0.01-0.033)E_{pl}$ for linear LIV and $(1.2-3.6)\times 10^{-13} E_{pl}$ for quadratic LIV for $s_{\pm}=1$~\cite{Piran06}.   For the Konus-Wind data (having a time resolution of 2~msec), spectral lags were calculated based on the 18-70 keV, 70-300 keV, and 300-1160 keV energy bands. No time delays between the different energy bands were detected, and  subsequently limits on LIV were set, whose values are given by $(0.0012-0.0066) E_{pl}$ for linear LIV and $(1.1-5.1) \times 10^{-13} E_{pl}$ for $s_{\pm}=1$~\cite{Piran06}. The slight variation in the aforementioned limits depends on which pair of energy bands were used to calculate the spectral lags. In this work, more stringent limits were obtained from Konus-WIND data due to its better time resolution.

A subsequent search was then done using Integral data~\cite{Lamon}. This work used unbinned maximum likelihood analysis of 11 GRBs with known redshifts to measure 17 time lags. A fit to the time delay consisting of  a constant intrinsic lag along with  LIV induced term was done. However the best-fit $\chi^2$/DOF was much greater than 1. After rescaling the error, a lower limit of $E_{QG}>3.2 \times 10^{11}$ GeV was found at 95\% c.l~\cite{Lamon}. This limit is valid for both subluminal and superluminal propagation.

Then a similar search was carried out using HETE-II data~\cite{Bolmont}. They considered 15 GRBs in the redshift range $z=0.16-3.37$ in the energy range of 6-400 keV. The spectral lags were calculated using de-noised light curves between 8-30 keV and 60-350 keV. A search for time lag as a function of both redshift and energy difference was carried out. No LIV signature was seen. A constant intrinsic time lag was considered and the data fit to Eq.~\ref{eq:sumdeltat}. A lower limit of $E_{QG}> 2 \times 10^{15}$ GeV was obtained at 95\% c.l. for $s_{\pm}=1$~\cite{Bolmont}.

About a month after the launch of the Fermi satellite, a bright GRB (GRB 080916C) located at a redshift (located at $z=4.35 \pm 0.15$) was detected~\cite{FermiScience}. Based on the spectral time lag of 16 seconds between a photon of energy 13 GeV and low energy photons of keV energy, a lower limit of $E_{QG}>1.2 \times 10^{19}$ GeV was obtained for $s_{\pm}=1$~\cite{FermiScience}.
Subsequently, a  LIV search was carried out using GRB 090510 located at $z=0.903 \pm 0.003 $ by calculating the spectral lag between photons above 30 GeV and below 1 MeV~\cite{Xiao09}. This gives a constraint of $E_{QG}>7.72 \times 10^{19}$ GeV and $E_{QG}>7.26 \times 10^{10}$ GeV for linear and quadratic LIV for $s_{\pm}=1$, respectively~\cite{Xiao09}. A similar analysis was also done by the Fermi LAT/GBM collaboration and multiple lower limits of $(10-100) E_{Pl}$ at 95\% c.l. were obtained for linear LIV for both values of $s_{\pm}$ (cf. Table 2 of ~\cite{Abdo}), where the limits change slightly depending on which energy bands were used to calculate the spectral lags~\cite{Abdo}. This was the first work which obtained a limit on $E_{QG}$ greater than the Planck scale for superluminal and subluminal LIV.

Another search for LIV was then done by considering 4 GRBs from Fermi-LAT spanning the redshift range from 0.9$-$4.13~\cite{Shao10}. Similar to ~\cite{Ellis}. they assumed that the spectral lag consists of a constant time lag along with a LIV-induced lag.
This work recast Eq.~\ref{eq:sumdeltat} and Eq.~\ref{eq:nullhypothesis} into a regression relation of the form (assuming $s_{\pm}=-1$)
\begin{equation}
\frac{\Delta t_{obs}}{1+z} = \Delta t_{int}  + K/E_{QG}^n, 
\label{eq:constt}
\end{equation}
where $K=\Delta t_{LIV} \times E_{QG}^n$. A separate fit was done for the three long GRBs as well as one  short GRB
in the sample. They showed that the spectral lag of the three long GRBs can be well fit by the sum of a constant lag along with a linear LIV lag~\cite{Shao10}. This was the first work using GRB spectral lags, which found  evidence for LIV    with $1\sigma$ bounded intervals for $E_{QG}$  given by: $E_{QG} = (2.2 \pm 0.9) \times 10^{17}$ GeV and 
$E_{QG} = (5.4 \pm 0.2) \times 10^{9}$ for linear and quadratic LIV, respectively~\cite{Shao10}. However they pointed out that if assumes that the observed lag has contributions from LIV, one obtains negative values for the intrinsic astrophysical lag, which is contrary to conventional wisdom for the astrophysical models, if low energy photons are produced by electrons and high energy ones by protons. Therefore, they urged that these hints for LIV should be treated with caution~\cite{Shao10}.

A different search for LIV   using Fermi-LAT GRBs was done by measuring the spacing between  photon bunches in four GRBs, which  were found to be  much shorter (for GRB 090510A) than any other temporal feature in the energy range from 1-30 GeV. ~\cite{Nemiroff11}. Photon bunching refers to more than three photons detected from GRB within a narrow time window (less than 0.01 seconds). Based on this, the most stringent bound on $\Delta c/c<6.94 \times 10^{-21}$ was obtained for GRB 090510A~\cite{Nemiroff11}.

Then, Vasileiou et al~\cite{Vasileiou13} considered 4 Fermi-LAT GRBs in the GeV energy range with measured redshifts. This work used three complementary techniques to quantify the amount of spectral dispersion in the data. Then using conservative choices for the source intrinsic spectral evolution, constraints were obtained on LIV.  For $s_{\pm}=1$, the most stringent limit was obtained for GRB 090510 with $E_{QG}>7.6 E_{pl}$ for linear and $E_{QG}>1.3 \times 10^{11}$ GeV for quadratic LIV~\cite{Vasileiou13}. For $s_{\pm}=-1$ the corresponding best limit was  $E_{QG}>18 E_{pl}$ using GRB 080916C for linear and $E_{QG}>6.7 \times 10^{10}$ GeV using GRB 090510  for  quadratic LIV~\cite{Vasileiou13}. 
The same group of  authors then analyzed Fermi-LAT  data from GRB 090510 to look for a stochastic spread in the speed of light around $c$~\cite{Vasileiou15}. No such variations were seen and 95\% c.l. lower limits on $E_{QG}$ are give by $E_{QG}>2.8 E_{pl}$~\cite{Vasileiou15}.


The next work which confirmed the previous tentative evidence for LIV found in ~\cite{Shao10},  was by ~\cite{Zhang}, which followed the same procedure as in ~\cite{Shao10,Ellis}. They analyzed the spectral lag data for 8 GRBs detected by both Fermi-LAT and Fermi-GBM.
They assumed a constant intrinsic lag in the source frame and fit the spectral lag to Eq.~\ref{eq:constt}.
They showed that the data from five long GRBs could be fit using a straight line with an intrinsic lag of $t_{int} = -12.1 \pm 1.7$ secs and $E_{QG} = (3.05 \pm 0.19) \times 10^{17}$ GeV for linear LIV hypothesis. After taking into account the uncertainty in the cosmological parameters and energy resolution of LAT, this uncertainty in  $E_{QG}$  is equal to   $0.7 \times 10^{17}$ GeV~\cite{Zhang}. If we consider a quadratic LIV, the best-fit value of $E_{QG}= (6.8 \pm 1.0) \times 10^9$ GeV~\cite{Zhang} with $s_{\pm}=-1$.  A variant of this analysis using a subset of the  GRBs  analyzed in ~\cite{Zhang} along with a few additional ones confirmed this tentative evidence for LIV  with  $E_{QG}=(3.6 \pm 0.26)\times 10^{17}$ GeV~\cite{Xu2}.
More confirming evidence was found using GRB 160509A (located at $z=1.17$) using photons in the energy range 1$-$50 GeV with $E_{QG}=3.6 \times 10^{17}$ GeV~\cite{Xu1}.  Two new methods for characterizing the arrival times of low energy photons needed for calculation of spectral lags in Fermi-GBM and Fermi-LAT data were proposed in ~\cite{Liu} and applied to data previously analyzed~\cite{Xu2,Xu1}. Although there was a slight difference in the regression relation, the results in ~\cite{Liu} were consistent with those in ~\cite{Xu2,Xu1}

The first search using a sample consisting  of only  short GRBs detected by SWIFT/BAT  and with redshifts in the range 0.36$-$2.2 was carried out in ~\cite{Bernardini17}. The rationale for selecting only short GRBs is that these GRBs have negligible spectral lag and smaller intrinsic dispersions compared to long GRBs. Therefore, choosing an exclusive sample of short GRBs would minimize any systematics related to the unknown nature of the intrinsic lag. 
The spectral lags were estimated between the 50-100 keV and 150-200 keV
energy intervals. This dataset of 15 GRBs was fit to a constant intrinsic lag and a LIV-induced lag (cf. Eq.~\ref{eq:constt}). While fitting the data, 
 an extra scatter was added to account for the dispersion in the spectral lag. However, unlike previous works~\cite{Shao10,Zhang,Xu1,Xu2}, the slope of the regression relation in Eq.~\ref{eq:constt} was consistent with zero, implying no LIV. Assuming a linear model for LIV, a 95\% c.l. lower limit on $E_{QG}$ was set given by $E_{QG}>1.48 \times 10^{16}$ GeV, valid for both values of $s_{\pm}$~\cite{Bernardini17}.

The first convincing case for a spectral lag transition  from positive to negative lags, using multiple lags over three decades in energy range from  the same  GRB  was demonstrated for GRB 160625B~\cite{Wei} located at $z=1.41$. This burst had three different sub-bursts   The spectral lags were calculated for 30 time lags starting from 12-16 keV to 17-20 MeV. This was also the first work, which did not assume a constant intrinsic lag model, and  instead used a phenomenological model for the intrinsic time lags given by:
\begin{equation} 
\Delta t_{int}^{rest} =\tau\Big[ \Big(\frac{E}{keV}\Big)^{\alpha}-\Big(\frac{E_0}{keV}\Big)^{\alpha}\Big],
\label{eq:delt_int}
\end{equation}
where $\alpha$ depicts the energy exponent and $\tau$ depicts the time scale for the intrinsic time lag, $E_0$ and $E$  correspond to the   lower and upper energy intervals. We note that if the energy interval corresponds to finite energy band, the midpoint of that band is usually used to ascribe an energy to it. This has been the norm in all literature on LIV searches using spectral lags. However, technically one should include the energy spectrum weighted values. This intrinsic model was empirically determined by modelling the single-pulse properties of about 50 GRBs~\cite{Shao}. This model has also been used for blazar flare modelling~\cite{Perennes20}.

 The analysis in ~\cite{Wei} found that the spectral lag for this  GRB  shows a turnover at 8 MeV, with  a transition from positive to negative lags. They argued that this transition could be a signature of LIV, which dominates at high energies. The best-fit value obtained for  $E_{QG}$    is given by:$\log(E_{QG}/GeV)=15.66^{+0.55}_{-0.01}$ and $\log(E_{QG}/GeV)=7.17^{+0.17}_{-0.02}$ for linear and quadratic LIV models, respectively assuming $s_{\pm}=1$~\cite{Wei}.  In a subsequent work, the statistical significance for this spectral lag transition was then  ascertained using both frequentist~\cite{Desai16b} and
 information theory techniques such as AIC and BIC~\cite{Liddle,Cosine,Krishak4,Dantuluri}. The frequentist test
  corresponds to a $Z$-score 
between $3.05-3.74\sigma$,  and  $\Delta$AIC/BIC $>10$ for the quadratic LIV model, which point to decisive evidence~\cite{Ganguly} (see also ~\cite{Gunapati})  However, despite finding a signature for LIV and obtaining marginalized closed contour intervals for $E_{QG}$, in the conclusions,  only
one-sided lower limits for $E_{QG}$ were reported in the conclusions given by, $E_{QG}>0.5 \times 10^{16}$ GeV  and $E_{QG}>1.4 \times 10^{7}$ GeV, for linear and quadratic LIV, respectively for $s_{\pm}=1$. Constraints on the   isotropic and anisotropic coefficients of the Lorentz violation SME
using this GRB were also obtained~\cite{Wei2}. However, subsequently it was  pointed out that the spectral lags are correlated with the spectral evolution in GRB pulses~\cite{Lu18}. This turnover in the spectral lag data could therefore be  the consequence of a  radiating relativistic jet shell with a cutoff BAND power law spectrum~\cite{Band93}, similar to that seen in observations~\cite{Du19}. This model was also used to explain the spectral lag transition in GRB 190530A, where however they used a three segment broken power law~\cite{Liang23}.

Subsequently,  the spectral lag data for GRB 160625B was stacked  together  with the data for 35 GRBs from ~\cite{Ellis} to look for LIV corresponding to $s_{\pm}=-1$~\cite{Pan}. Using the same intrinsic model as ~\cite{Wei} this stacking  procedure provides a more robust estimate of the intrinsic time lag.  The best-fit estimates  obtained for  $\log(E_{QG}/GeV)$ are equal to $14.523^{+0.022}_{-0.025}$ and $8.79 \pm 0.0097$ for linear and quadratic LIV models, respectively~\cite{Pan}.   This was also the first work which used a model-independent estimate of the expansion history using Gaussian process regression~\cite{Seikel_2012,Haveesh} based on $H(z)$ measurements from cosmic chronometers~\cite{Jimenez_2002}.

Most recently, a similar spectral lag transition from positive to negative lags (similar to GRB 1606025B)  was also detected in GRB 1901114C (detected by SWIFT) at about 0.7 MeV~\cite{Du}. This work calculated the spectral lags in energy intervals from 15-5000 keV and compared them to the lowest energy value of 12.5 keV.
However, the model for the intrinsic time lag asserted  was opposite in sign compared to ~\cite{Wei,Pan}. The statistical significance of the spectral lag transition was also estimated using Bayesian model comparison and found to be $>100$, pointing to decisive evidence using Jeffrey's scale. Similar to ~\cite{Wei,Pan}, they obtained best-fit values for linear and quadratic models, with  their 2$\sigma$ bounds  given by $\log(E_{QG}/GeV)$= $14.49^{+0.12}_{-0.13}$ and $6.00 \pm 0.06$, respectively assuming $s_{\pm}=1$.   They showed using $\chi^2$/DOF that all the LIV models  provide a good fit to the LIV hypothesis~\cite{Du}. 

The MAGIC collaboration however failed to find a similar evidence for an energy-dependent speed of light in the TeV gamma ray data for the same GRB. Using conservative assumptions on spectral and temporal evolution, the MAGIC collaboration  set a 95\% c.l. lower limit on $E_{QG}$ of  $0.55 \times 10^{19}$ GeV and $5.6 \times 10^{10}$ GeV for linear and quadratic models, respectively for superluminal propagation~\cite{MAGIC}. For subluminal propagation, the corresponding limits are $0.28 \times 10^{19}$ GeV and $7.3 \times 10^{10}$ GeV for linear and quadratic models, respectively~\cite{MAGIC}.

Therefore, the situation as of 2021 was that  some of  the above aforementioned analyses  obtained closed $1\sigma$ confidence intervals for $E_{QG}$, which is less than the Planck energy scale (indicating evidence for LIV), although not with the same value of $s_{\pm}$. However,  the estimated LIV energy scale estimated in ~\cite{Wei,Pan,Du} contradicts the previous  most stringent lower limits on LIV~\cite{Abdo,Vasileiou13,Vasileiou15}. Still, given the tantalizing hints for LIV in some  of the aforementioned works~\cite{Ellis,Wei,Pan,Du}, if the spectral lag data for all the GRBs can be described by a unified model, one would expect the net statistical significance of the LIV  to get  enhanced, if one stacks the spectral lag data from  all the GRBs, and analyze them uniformly with the same model. Therefore, in order to resolve this imbroglio, the   data from all the  works which found these hints for LIV signatures in the spectral lag data~\cite{Ellis,Wei,Pan,Du}  were stacked together~\cite{Agrawal21}. In all, there were a total of 91 spectral lag  measurements from 37 GRBs, including 19 lags from GRB 190114C and 37 lags from GRB 1606025B.
The aforementioned work did  a Bayesian model comparison  that the combined spectral lag data is a combination of both an intrinsic and a  LIV-induced lag, as opposed to only an intrinsic lag. The model assumed for the intrinsic astrophysical emission is the same as that proposed in ~\cite{Wei,Pan}. The priors used for   $\alpha$ and $\tau$ were broad enough to encompass both positive and negative values. The expansion history was estimated in a non-parametric fashion using Gaussian process regression.
When searching for LIV, no closed contour was obtained  for linear LIV (see Fig. 2 of ~\cite{Agrawal21}). Only the quadratic LIV showed a closed contour. They found that the Bayes factor for both the LIV models was greater than 100 relative to the null hypothesis.  The 68\% c.l. lower limit for $E_{QG}$ was $\log [E_{QG} (GeV)] > 16.08$ for linear LIV, whereas for quadratic LIV the best-fit value obtained was $\log [E_{QG} (GeV)] = 7.17 \pm 0.07$ for quadratic LIV~\cite{Agrawal21}. All these limits assumed $s_{\pm}=1$. However, the $\chi^2$/DOF  was still very much greater than one when all the three datasets were combined. Various other variants were tried in this work (such as a constant lag in conjunction with intrinsic scatter, two different exponents in the intrinsic emission for GRB 190114C), but the reduced $\chi^2$ was still much greater than one for these other combinations also.

Another  search for LIV (for $s_{\pm}=-1$) using eight Fermi-LAT GRBs with energies above 100 MeV located at redshifts between 0.34 and 4.35 has also been recently carried out~\cite{Ellis19}. This work considered
three distinct non-parametric measures based on irregularity, skewness, and kurtosis  to characterize the GRB emission. A constant  lag at the source frame was used for the intrinsic emission. This work obtained a 95\% lower limit of $E_{QG}>8.4 \times 10^{17}$ GeV~\cite{Ellis19}. If one takes into account the intrinsic temporal spectral variations, the 95\% lower limit  becomes $E_{QG} > 2.4\times 10^{17}$ GeV~\cite{Ellis19}.

A novel search for LIV using a forward modelling technique, (which involves constructing  individual  Monte Carlo models for every source,  obtained by convolution of LIV as well as Gaussian Mixture Model based likelihood, which incorporates the redshift uncertainties) was carried out using the BATSE GRB catalog~\cite{Bartlett}. This work used the GRB catalog compiled in ~\cite{Yu18} after culling the GRBs with pseudo-redshifts greater than 4.5. They considered two LIV theories. In the first one, the photon speed receives quadratic corrections due to the photon mass. In the second scenario, the photon speed depends linearly on the energy (similar to the $n=1$ limit in Eq.~\ref{eq:vE}). The expected time delays of photons were obtained using a novel forward modelling approach. This work obtained a limit on photon mass $m_{\gamma}< 4.0 \times 10^{-5} h$ eV
and a quantum gravity length scale $l_{QG}< 5.3 \times 10^{-18} GeV^{-1}$ at 95\% c.l. (where $l_{QG} \equiv E_{QG}$~\cite{Bartlett}.  These results are also agnostic to the astrophysical and observational uncertainties related to the spectral lag. These limits are also valid for both subluminal and superluminal propagation, where $l_{QG}$ is negative.

Most recently, a comprehensive search using 135 long GRBs from the 2020  Fermi/GBM catalog  was carried out~\cite{Liu22}. Evidence for  spectral lag transitions (from positive to negative) have been found for  32 of these GRBs in the redshift range 0.347-4.35~\cite{Liu22}. For each GRB, close to 20  spectral lags between 10 keV to 10,000 keV (cf. Table 1 of ~\cite{Liu22}) were used to determine the transition.  A broken power law was used to model the intrinsic spectral lag, which is given by:
\begin{equation}
\Delta_{int} = \zeta \left(\frac{E-E_0}{E_b}\right)^{\alpha_1}\left\{0.5\left[1+\left(\frac{E-E_0}{E_b}\right)^{1/\mu}\right]\right\},
\label{eq:bkp}
\end{equation}
where $\zeta$ denotes the normalization amplitude, $\alpha_1$ and $\alpha_2$ represent the pre- and post-transition slopes before and after the transition energy $E_b$ and $\mu$ characterize the smoothness of the transition. Eq.~\ref{eq:bkp} reduces to Eq.~\ref{eq:delt_int} when $\alpha_1=\alpha_2$.  Evidence  for spectral lag transitions (from positive to negative) have been found using 32 Fermi/GBM GRBs in the redshift range 0.347-4.35~\cite{Liu22}.
The results from LIV search from {\it each} of  the  32 GRBs for $s_{\pm}=1$  can be found in Table 3 of ~\cite{Liu22}. The statistical significance of evidence for LIV based on the observed spectral lag transitions however has not been carried out for  these GRBs. However, the authors have argued that some of the negative lags are astrophysical. For the linear LIV model, the lower limits for $E_{QG}$ range from $(8.2 \times 10^{12} - 5.5  \times 10^{15})$ GeV~\cite{Liu22}. For the quadratic LIV model, the corresponding limits for $E_{QG}$ range from ($6.2 \times 10^4 - 1.7 \times 10^7$ GeV)~\cite{Liu22}. Limits on the  anisotropic LIV coefficients from SME were also set using this data~\cite{Jinnanwei22}.

Finally, observations of two potentially lensed GRBs (GRB 950830 and GRB 200716C) have also been used to search for LIV~\cite{LinLan2022}. Since lensing is achromatic, the lensed images should have the same hardness and magnification. However, different hardness values were observed in the energy bands, which could be induced due to  LIV as argued in ~\cite{LinLan2022}. The LIV induced time delays due to lensing were then calculated and is energy dependent. For GRB 950830, the lens  was estimated to have a redshift $z \sim 1$ and  the source at redshift of about two. For GRB 200716C, the lens redshift was estimated to be $z \sim 0.174$ and the source at $z \sim 0.348$. The observed time delays were then used to constrain LIV  assuming a point source lens mass. The best-fit values for a linear model of LIV and  $s_{\pm}=1$  is given by $E_{QG} \geq 3.2 \times 10^9$ GeV  (GRB 950830) and $E_{QG} \geq 6.9 \times 10^9$ GeV (GRB 200716C)~\cite{LinLan2022}.

\subsection{Searches using fixed source frame energies}
All the above studies have been done based on the spectral lags in a fixed energy interval in the observer frame.  However, if the  data contains GRBs with different redshifts, the corresponding energy interval in the source frame would  be  different. It has been  shown that is  a large scatter in the correlation between the observer-frame lags and the source-frame lags for the same GRB sample, implying that the observer-frame lags do not  represent the rest-frame lag~\cite{Ukwatta2012}.  Therefore, the observer-frame lags would be strongly biased, since  they  sample different parts of the intrinsic light curves. Therefore, this could introduce systematics in the search for LIV~\cite{Wei17,Wei2021}. To ameliorate this, the search for LIV was done by calculating the spectral lags between two fixed energy intervals in the intrinsic or {\it source} frame~\cite{Wei17}. Therefore, Eq.~\ref{eq:deltaliv} and Eq.~\ref{eq:constt} get modified  for a linear LIV and $s_{\pm}=-1$ as follows~\cite{Wei17}:
\begin{equation}
\Delta t_{LIV} =  \frac{(E_h' - E_l')}{E_{QG} H_0 (1+z)}\int_{0}^{z} \frac{(1+z^{\prime})}{h(z^{\prime})} \, dz^{\prime},
\label{eq:restframe}
\end{equation}
where $E_h'$ and  $E_l'$ are the rest frame energies in the upper and lower energy intervals, respectively. For this analysis, a dataset of 56 rest-frame spectral lags from the SWIFT satellite compiled in ~\cite{Bernardini} was used~\cite{Wei17}. This work also assumed a constant intrinsic lag (in the source frame). The spectral lags were again fit to a linear regression (similar to ~\cite{Ellis} and other aforementioned works) with the  Lorentz violating term corresponding to the slope and the intrinsic delay corresponding to the intercept. The best-fit value for slope was consistent with 0 to within $1\sigma$, implying there is no evidence for LIV. The 95\% c.l. lower limit for $E_{QG}$ is given by $E_{QG} > (2-2.2) \times 10^{14}$ GeV for $s_{\pm}=-1$~\cite{Wei17}. This was also the first work which divided the GRB sample into four groups according to  redshift to look  for a redshift-dependent trend in the LIV coefficients. No such variation was seen after splitting the data. An independent search for  evidence for LIV using this dataset  for $s_{\pm}=1$ and using  the same intrinsic emission model as Eq.~\ref{eq:delt_int} was also carried out~\cite{Agrawal21}. No evidence for either of the two  LIV  models was found  using Bayesian  model comparison tests~\cite{Agrawal21}.

Most recently  a search for LIV using a catalog of  spectral lags from 44 short GRBs and 2 long GRBs from both SWIFT and FERMI-GBM was carried out~\cite{Xiao22}. The lags were constructed between the  fixed source frame energy intervals of 15-70 keV and 120-250 keV, and were obtained using  a novel cross-correlation method, which utilizes the temporal information in the light curve and is agnostic to the details of the cross-correlation function~\cite{Li04,Xiao21}.   A constant
source frame intrinsic lag was used. Both AIC and BIC were used to assess  the significance of LIV.  The intrinsic constant lag model as well as the LIV models provide equally good fits to the data.  From this analysis,  a lower limit on $E_{QG}$ in the range $10^{15}-10^{17}$ GeV was obtained at 95\% c.l., depending on the choice of $s_{\pm}$~\cite{Xiao22}. A refined analysis was also carried out using the energy-dependent intrinsic lag (Eq.~\ref{eq:delt_int}),  chronometers to characterize the expansion history, and also incorporating the uncertainties in the energy intervals with these changes~\cite{Desai23}. Unlike ~\cite{Xiao22}, no closed contours were obtained for $E_{QG}$ for both the linear and quadratic LIV hypotheses. The 95\% c.l. lower  limits obtained are $E_{QG}>4 \times 10^{15}$ GeV and $E_{QG}>6.8 \times 10^9$ GeV, respectively for $s_{\pm}=1$~\cite{Desai23}.

In addition to LIV with a pure sample of GRBs or a pure sample of AGNs, a search for LIV using a combined sample of GRBs and AGNs has also been done using a combination of 24 GRBs and AGNs detected by Fermi-LAT, RHESSI, and TeV detectors (such as HESS, MAGIC and WHIPPLE)~\cite{Kislat}. This data was used to constrain 25 coefficients of the anisotropic birefringent Lorentz violating SME~\cite{Kislat}.

\subsection{Searches based on  TeV-PeV observations}
All the previous results for LIV  were based on spectral lag measurements for GRBs  in the  keV-GeV energy regime (except for the MAGIC result which used TeV observations of one GRB). These works have used calculations of  the opacity of the universe to very high energy gamma-rays in presence of LIV, such as ~\cite{Abdalla18}.
Here, we discuss LIV searches based on one GRB detected at energies above 100 TeV, which has generated a lot of interest and excitement.

In October 2022, GRB 221009A  was detected at energies at 18  TeV by the LHAASO observatory~\cite{LHAASO} and upto 250 TeV by the Carpet-2 detector~\cite{Carpet2}. This burst was also seen by Fermi-LAT in the 0.1-1.0 GeV range. This burst is also the brightest in terms of its peak flux and fluence~\cite{Burns23}. The lower limit  on the flux  observed at LHAASO  based on the detection of 1 photon is $\frac{dN}{dE} \geq 2.9 \times 10^{-19}~\rm{ph~cm^{-2}~s^{-1} ~GeV^{-1}}$. Extrapolating the Fermi-LAT spectrum to this energy range gives a flux of $9.3 \times 10^{-12}~\rm{ph~cm^{-2}~ s^{-1}~GeV^{-1}}$ . In a similar vein, the lower limit on the flux detected by Carpet-2 at 250 TeV is given by  $\frac{dN}{dE} \geq 1.8 \times 10^{-16}~\rm{ph~cm^{-2}~ s^{-1}~GeV^{-1}}$. The corresponding value extrapolated from Fermi-LAT is given by $6.7 \times 10^{-14}~\rm{ph~cm^{-2}~s^{-1} ~GeV^{-1}}$. Based on a comparison of the observed a interpolated spectrum at 18 TeV, a 95\% c.l. upper limit on the opacity ($\tau$) , given by $\tau \leq 17$ was obtained~\cite{Finke23}. This constraint is consistent with only a small fraction  of models of extragalactic background light~\cite{Baktash22}. The corresponding limit at 250 TeV is given by $\tau \leq 5.9$, which is in severe tension with most models of extragalactic background light. This gives a {\it upper}  limit on $E_{QG}$, which is given by $E_{QG} \leq 49 E_{pl}$ (for linear LIV) and  $E_{QG} \leq 10^{-6} E_{pl}$ for quadratic LIV. Therefore, taken at face value,  this also constitutes evidence for LIV. Other works that have analyzed this GRB have also come to similar conclusions~\cite{Baktash22,Zhang22,Galanti22,LiMa23}. Again this result is in conflict with the lower limits obtained in ~\cite{Vasileiou13}. However,  this result also has a number of caveats associated with it as discussed in ~\cite{Finke23}, from prosaic astrophysical assumptions~\cite{Sahu23}  to other exotic Physics, which could explain the disappearance of photons~\cite{Galantiaxion}. This result should soon be confirmed with more detections from LHAASO, HAWC,  and the upcoming Cherenkov Telescope Array, which could detect VHE emission from GRBs upto 100s of TeV.

\section{Searches using polarization observations}
\label{sec:polarization}
In addition to searches for LIV with photon counts, corresponding searches have also been carried out from polarization observations using GRBs as well as with AGNs. We now recap some of these results.

\subsection{Searches with GRBs}
The first test for LIV using polarization measurements was carried out for GRB  041219A located at a redshift of $z= 0.02$~\cite{Laurent11}.  In the presence of LIV, the photon dispersion relation for frequency ($\omega$) and wavevector ($k$) gets modified to the following form:
\begin{equation}
 \omega^2=k^2 \pm \frac{2 \xi k^3}{M_{pl}} , 
\end{equation}
where $\xi$ is the LIV term and the rotation of direction of polarization after traversing a distance $d$ is given by:
\begin{equation}
\Delta \theta (p) \approx \xi \frac{k^2 d}{2 M_{pl}} .
\end{equation}
In terms of cosmological parameters this can be written as:
\begin{equation}
\Delta \theta (p) = \xi \frac{k^2 }{M_{pl} H_0} \int_{0}^{z} \frac{(1+z^{\prime})}{h(z^{\prime})} \, dz^{\prime},  
\end{equation}

The two different values of $\omega$ cause a rotation of the polarization   during the propagation of linearly polarized photons.
For this GRB a large amount of polarization was observed~\cite{McGlynn}. A search for a polarimetric angle shift between 200-250 keV and 250-325 keV, (which could be signature of LIV) was carried out. No such shift was seen. Based on this a constraint on the coupling  $\xi$ of the dimension five  Lorentz violating  interaction was obtained, which is given by $\xi<10^{-14}$~\cite{Laurent11}.

The subsequent search for LIV using polarized emission was done using GRB 110721A, GRB 100826A and GRB 110301A using polarized emission from the IKAROS satellite~\cite{Toma12}. The polarized emission was detected in two separate energy  bands of 70-100 keV and 100-300 keV. A pseudo-redshift was estimated based on using the tightness of the $E_{peak}-L_{peak}$ correlation. For constraints on LIV  the $2\sigma$ lower limit on the redshift was used.  The most stringent limit was obtained for GRB 110721A with $|\xi|<2 \times 10^{-15}- 8 \times 10^{-16}$, depending on the assumption used for the  amount of intrinsic polarization~\cite{Toma12}.

Subsequently, a search for LIV  using  polarization measurements obtained by ASTROSAT for seven GRBs~\cite{Astrosat}, as well as all of the five  previously detected polarized GRBs was carried out~\cite{Weipolarization} by applying a new technique proposed in ~\cite{Lin16}.  For GRBs with unconfirmed redshifts, lower limits on each of the GRB redshifts were obtained by assuming the GRBs fall within $2\sigma$ of the Amati relation~\cite{Amati02,Govindaraj}. This work pointed out that all previous LIV-based limits from polarized GRBs, which were obtained by assuming  that the rotation angle of the polarization vector ($\Delta \theta$), given by  $\Delta \theta \leq \pi/2$ could be ameliorated by applying the technique proposed in ~\cite{Lin16}. Subsequently new limits were set using each of the 12 GRBs (cf. Table 1 in ~\cite{Weipolarization}) with the most stringent limit given by $\xi<5.2 \times 10^{-17}$ for GRB 061122~\cite{Weipolarization}

In addition to polarization from the prompt emission phase of GRBs in the high energy range, searches for LIV from using polarization observations in the afterglow phase have also been carried out~\cite{WeiWu20}. This work used optical afterglow observations of GRB 020813 and GRB 021004 to look for birefringence. For GRB 020813, the best-fit value of $\xi$ was $\xi = 1.58 \times 10^{-7}$ and is consistent with a zero value to within $2.5\sigma$~\cite{WeiWu20}. The corresponding value for GRB 021004 was $\xi=-0.1^{+1.2}_{-1.7}$ at the $3\sigma$ level~\cite{WeiWu20}.

More stringent limits on $\xi$ can be obtained from upcoming missions such as POLAR, TSUBAME, COSI, and GRAPE, which will carry out polarimetry of the prompt emission phase of the GRBs~\cite{McConnell}. Other possible avenues for searching for LIV using astrophysical polarization measurements are reviews in ~\cite{Kislat}.

\subsection{Searches with AGNs}
A search for LIV-induced birefringence  caused by rotation in the polarization vector of a linearly polarized plane wave has also been done using optical spectropolarimetry from five blazars (3C 66A, S5 0716+714, OJ 287, MK 421, and PKS 2155-304)~\cite{Zhou21}.
Polarization measurements for these aforementioned blazars were carried out using 37 groups of observations in five optical filters $UBVRI$. The $2\sigma$ limits on the five blazars are given by $-8.91 \times 10^{-7} < \xi < 2.93 \times 10^{-5}$~\cite{Zhou21}.

A similar proof of principles test using observations of two  AGNs (BL Lacartae and S5 B0716+714) was also carried out with the Array Photo polarimeter and constraints on Lorentz violation SME were obtained based on these observations~\cite{Friedman}.


\begin{table}[h]
\begin{tabular}{l|c|c|c|c}
\hline
\hline
\textbf{Method} & \textbf{Source} &  \textbf{$E_{QG}$ (Linear LIV)} & \textbf{$E_{QG}$ (Quadratic LIV)} & \textbf{Reference} \\
& & (GeV) & (GeV) & \\ \hline
Spectral Lag & Pulsar (Crab) & $>10^{25}$  & - & \cite{LHAASOCrab} \\
Spectral Lag & AGN (Stacked) & $>1.28 \times 10^{20}$ & $>2.38 \times 10^{12}$ & \cite{Lang22} \\
 Relative Spectral Lag & Neutrinos-GRB   & $>8 \times 10^{17}$ & - & \cite{ZhangYang22} \\
Spectral Lag & GRB 090510 & $>1.02 \times 10^{21}$ & $>1.3 \times 10^{11}$ & \cite{Abdo,Vasileiou13}  \\
$\gamma$-ray propagation & 56 stacked sources &  $>12.08 \times 10^{19}$ & $>2.38 \times 10^{12}$ & \cite{Lang19} \\
TeV - PeV observations   & GRB 221009A & $\leq 49 \times 10^{19}$ & $\leq 10^{13}$ & \cite{Finke23} \\
\hline
\end{tabular}
\caption{\label{tab:table}Summary of the most stringent 95\% c.l. limits on $E_{QG}$ for linear as well as quadratic LIV, using  various astrophysical sources  discussed in this review. An exhaustive tabular summary of other LIV observables ($\Delta c/c$, $\xi$, LIV SME coefficients, etc.) from literature can be found in  ~\cite{WeiWu21}.}
\end{table}
\section{Conclusions}
\label{sec:conclusions}
In this manuscript, we have reviewed observational tests of LIV from neutrinos (using terrestrial and astrophysical observations), gravitational waves, photon based astrophysical observations of GRBs, AGNs, and pulsars, and cosmological tests with galaxy clusters, BAO, Type Ia SNe and CMB. Most of the tests using GRBs and AGNs have been carried out using the observed spectral lags (in the keV-GeV energy range) by looking for an energy-dependent speed of light. Tests with polarization measurements  entail looking for a change in polarization angle due to vaccuum birefringence.
Among these, some works have found evidence for such an energy dependence based on fitting the spectral lags, although not for the same value of $s_{\pm}$~\cite{Shao10,Zhang,Xu1,Xu2,Wei,Du,Pan,Xiao22}. Evidence for LIV based on  TeV-PeV observations of  GRB 221009A  has also been claimed (although with caveats)~\cite{Finke23}. However, the inferred values of $E_{QG}$  are in conflict some of the most stringent limits obtained~\cite{Vasileiou13,Vasileiou15,Lang19}. A summary of the most stringent lower limits (and also one upper limit) on $E_{QG}$ from the different types of searches are summarized in Table~\ref{tab:table}. (See also Table 3 of ~\cite{WeiWu21} for a comprehensive summary of other LIV observables). 
Furthermore, when one combines the data from  these disparate datasets, which individually point to LIV, one does not get a consistent picture or a good fit to the combined data~\cite{Agrawal21}. Also, it is not easy to disentangle the LIV-induced lags from astrophysical contributions to the lag, which is still unknown. Many of the aforementioned works have used different models for the astrophysical lag.
Given the diversity in the observed GRB light curves, one would need to incorporate these  observational  based priors,  while modelling the intrinsic spectral lags in order to make a robust case for evidence of LIV. The recent observation of GRB 221009A at energies above 100 TeV is exciting. More observations of GRBs at such high energies should soon be forthcoming from high energy gamma ray telescopes at energies above  $>$ 100 GeV~\cite{Zhang19}.

In the area of cosmology the recent evidence for birefringence is tantalizing, and it would be interesting to see how this pans out with additional data from upcoming CMB missions~\cite{Komatsu22}.

%

\begin{acknowledgement}
 I would like to thank Cosimo Bambi and  Alejandro C\'ardenas-Avenda\~no for invitation to write this chapter as well as for providing detailed comments on the manuscript. Finally I acknowledge  Shalini Ganguly, Haveesh Singirikonda, and Rajdeep Agrawal for prior collaboration in this area.
\end{acknowledgement}

\biblstarthook{
}

\bibliography{chapter}

\end{document}